\documentclass[reprint,amsmath,amssymb,aps,superscriptaddress,longbibliography]{revtex4-2}
\usepackage{graphicx}
\usepackage{dcolumn}
\usepackage{bm}
\usepackage{color}
\usepackage[dvipsnames]{xcolor}
\usepackage[bookmarks=true,colorlinks,linkcolor=OrangeRed,urlcolor=NavyBlue,citecolor=RoyalBlue]{hyperref}
\usepackage{braket}
\hypersetup{nolinks=true}

\usepackage{soul} 

\AtBeginDocument{\let\oldcontentsline\contentsline}
\newcommand{\notoccontentsline}[4]{\oldcontentsline{}{}{}{}}
\newcommand{\droptocpage}{\addtocontents{toc}{\let\protect\contentsline\protect\notoccontentsline}}
\newcommand{\incltocpage}{\addtocontents{toc}{\let\protect\contentsline\protect\oldcontentsline}}

\begin{document}

\title{Probing false vacuum decay on a cold-atom gauge-theory quantum simulator}

\author{Zi-Hang Zhu}
\affiliation{Hefei National Research Center for Physical Sciences at the Microscale and School of Physical Sciences, University of Science and Technology of China, Hefei 230026, China}

\author{Ying Liu}
\affiliation{Hefei National Research Center for Physical Sciences at the Microscale and School of Physical Sciences, University of Science and Technology of China, Hefei 230026, China}

\author{Gianluca Lagnese}
\affiliation{Jo\v zef Stefan Institute, Jamova cesta 39, 1000 Ljubljana, Slovenia}

\author{Federica Maria Surace}
\affiliation{Department of Physics and Institute for Quantum Information and Matter,
California Institute of Technology, Pasadena, California 91125, USA}

\author{Wei-Yong Zhang}
\affiliation{Hefei National Research Center for Physical Sciences at the Microscale and School of Physical Sciences, University of Science and Technology of China, Hefei 230026, China}

\author{Ming-Gen He}
\affiliation{Hefei National Research Center for Physical Sciences at the Microscale and School of Physical Sciences, University of Science and Technology of China, Hefei 230026, China}

\author{Jad C.~Halimeh}
\affiliation{Max Planck Institute of Quantum Optics, 85748 Garching, Germany}
\affiliation{Department of Physics and Arnold Sommerfeld Center for Theoretical Physics (ASC), Ludwig Maximilian University of Munich, 80333 Munich, Germany}
\affiliation{Munich Center for Quantum Science and Technology (MCQST), 80799 Munich, Germany}

\author{Marcello Dalmonte}
\affiliation{The Abdus Salam International Centre for Theoretical Physics (ICTP), Strada Costiera 11, Trieste 34151, Italy}

\author{Siddhardh C.~Morampudi}
\affiliation{Center for Theoretical Physics, Massachusetts Institute of Technology, Cambridge, MA 02139, USA}

\author{Frank Wilczek}
\affiliation{Center for Theoretical Physics, Massachusetts Institute of Technology, Cambridge, MA 02139, USA}
\affiliation{T. D. Lee Institute and Wilczek Quantum Center, SJTU, Shanghai}

\author{Zhen-Sheng Yuan}
\affiliation{Hefei National Research Center for Physical Sciences at the Microscale and School of Physical Sciences, University of Science and Technology of China, Hefei 230026, China}
\affiliation{Hefei National Laboratory, University of Science and Technology of China, Hefei 230088, China}

\author{Jian-Wei Pan}
\affiliation{Hefei National Research Center for Physical Sciences at the Microscale and School of Physical Sciences, University of Science and Technology of China, Hefei 230026, China}
\affiliation{Hefei National Laboratory, University of Science and Technology of China, Hefei 230088, China}

\begin{abstract}

In the context of quantum electrodynamics, the decay of false vacuum leads to the production of electron-positron pair, a phenomenon known as the Schwinger effect. In practical experimental scenarios, producing a pair requires an extremely strong electric field, thus suppressing the production rate and making this process very challenging to observe. Here we report an experimental investigation, in a cold-atom quantum simulator, of the effect of the background field on pair production from the infinite-mass vacuum in a $1+1$D $\mathrm{U}(1)$ lattice gauge theory. The ability to tune the background field allows us to study pair production in a large production rate regime. Furthermore, we find that the energy spectrum of the time-evolved observables in the zero mass limit displays excitation peaks analogous to bosonic modes in the Schwinger model. Our work opens the door to quantum-simulation experiments that can controllably tune the production of pairs and manipulate their far-from-equilibrium dynamics.
\end{abstract}

\date{\today}
\maketitle
\droptocpage

\textbf{\textit{Introduction.---}}
The instability of the quantum electrodynamic (QED) vacuum with respect to the creation of electron--positron pairs in the presence of a strong external background electric field was predicted almost 70 years ago and has not been directly observed yet \cite{sauter1931behavior,heisenberg1936folgerungen,PhysRev.82.664,AFFLECK198238,AFFLECK1982509}. 
Because of quantum fluctuations, electron--positron pairs are continuously created and annihilated through virtual (off-shell) processes --- also known as \textit{vacuum polarization} --- and they cannot be observed. 
In the presence of a sufficiently strong external electric field, a pair can acquire enough energy from the field such that the energy necessary to create the particle is compensated, and the pair becomes real (on-shell). 
This phenomenon is referred to as the \textit{Schwinger effect}~\cite{PhysRev.82.664}, which has a mesoscopic variant called the Zener breakdown effect describing the generation of electron--hole pairs in condensed matter~\cite{schmitt2023mesoscopic}. 
However, for the production of electron--positron pairs, when the electric field intensity $E$ is much smaller than the so-called \textit{Schwinger limit}, which is about $E_c \approx 10^{16}$ V/cm, pair production is exponentially suppressed in $1/E$ (see~\cite{GELIS20161} for a review). 
Such dependence, indicative of the non-perturbative nature of the phenomenon, makes experimental observations extremely hard.

\begin{figure}[!htb]
    \includegraphics[width=\linewidth]{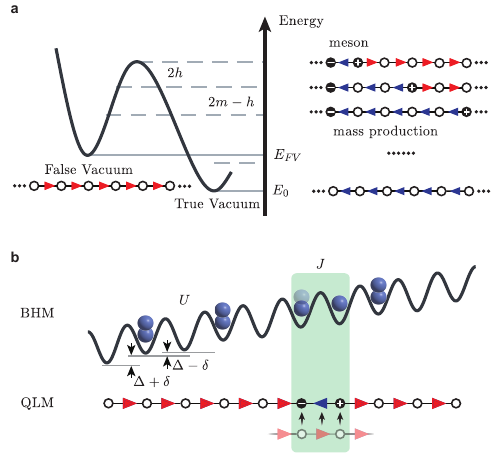}
    \caption{ 
    \textbf{Depiction of false vacuum decay and implementation of target $\mathrm{U}(1)$ quantum link model}.
    (a) Illustration of false decay. 
    The particle pair is created virtually from a false vacuum state while energy increases by $2m-h$, then expands to reduce the energy. 
    Finally, the initial false vacuum state decays into a bubble of true vacuum, which lies between the particle pair, such that the final state has the same energy as the initial state (resonant bubble).
    (b) Schematic optical superlattice for bosons in a linearly tilted potential. $U$ is the on-site interaction strength, $J$ is the hopping amplitude of bosons, $\delta$ is the energy offset between neighboring shallow and deep lattice sites, and $\Delta$ is the linear tilt per site. 
    In the mapping, each pair of adjacent sites is interpreted as a gauge site, where the left (right) arrow indicates the direction of the electric field. 
    After determining the gauge link configuration, charges in matter sites are obtained by Gauss's law, where circles with $+(-)$ indicate the positive (negative) charge and hollow circles represent the vacuum. 
    The hopping process between $|..20..\rangle \leftrightarrow |..11..\rangle$ in the Bose--Hubbard model corresponds to the matter-gauge coupling in QLM. 
    When we tune $\delta$ to a non-zero value, two $\mathbb{Z}_2$ vacuum states are no longer degenerate. 
    This is equivalent to introducing a background electric field $h$. 
    }
    \label{fig1}
\end{figure}

This spontaneous pair production can be understood as the decay of a metastable state (false vacuum) \cite{PhysRevD.15.2929,PhysRevD.16.1762,PhysRevD.21.3305,turner1982our}: the particle--antiparticle pair has to tunnel across a large distance (inversely proportional to the electric field) to compensate for the energy needed to produce the pair, making this tunneling extremely unlikely (Fig.~\ref{fig1}(a)). 
Similar phenomena of false vacuum decay are of fundamental interest in a wide range of physical systems \cite{LANGER1967108,langer1969statistical,Affleck1981,debenedetti2001supercooled,brooke2001tunable,budden2021evidence,miyashita2022collapse}.
For example, understanding the decay process is expected to help elucidate the structure of the Higgs vacuum in particle physics, as well as the inflation of the early universe in cosmology~\cite{elias2012higgs,PhysRevD.27.2848,fialko2015fate,ng2021fate}. 
However, the inherent exponential complexity of this phenomenon raises significant challenges for both experimental and theoretical investigations. 
In many circumstances, the long lifetime of a false vacuum state makes it virtually impossible for us to observe the decay process directly.

A promising setting to explore such phenomena is provided by quantum simulators and quantum computers, on which quantum statistical mechanics models can be efficiently realized. 
Prominent examples in this direction have been the study of metastable states in quantum spin systems and Bose gases~\cite{Billam2019,PhysRevB.102.041118,PhysRevB.104.L201106,Abel2021,Billam2022,Song2022}, and the recent observation of bubble formation in Bose-Einstein condensates~\cite{zenesini2023observation}. 
Further, the recent success in implementing lattice gauge theories in synthetic quantum matter ~\cite{Martinez2016,Schweizer2019,Chin2019,Mil2020,Yang2020,Zhou2022,PRXQuantum.3.020303,PhysRevResearch.4.L022060,PhysRevResearch.5.023010,PhysRevLett.131.050401,nguyen2022digital} represents an unprecedented opportunity to study false vacuum decay, in a context where controlled laboratory dynamics can be realized in the presence of a key aspect of field theories --- gauge symmetry.

Here, we report the direct observation of pair creation from the false vacuum in a $1+1$D $\mathrm{U}(1)$ lattice gauge theory --- in its quantum link model formulation --- using a cold-atom quantum simulator. 
Our experiment utilizes staggered superlattices and programmable potential engineering to realize gauge field dynamics with a tunable background electric field (a $\theta$ angle) \cite{PRXQuantum.3.040316,PRXQuantum.3.040317}, providing a controllable platform for studying pair production. 
By tuning the strengths of this background electric field, we can tune the instability of the initial state: we can simulate a regime where the pairs can be resonantly produced at a short distance, enhancing their production rate. 
We then monitor the system's dynamics with atom-number-resolved quantum gas microscopy. 
We observe that the production rate of the particle--antiparticle pairs is suppressed either as the fermion mass is increased or when the external field is dominant. 
Moreover, as the mass approaches infinity, we find that the time needed to reach maximum pair production agrees well with a model of dilute (decoupled) particle--antiparticle pairs~\cite{PhysRevX.6.011023}. 
Thanks to direct access to the system wave function, along with the particle production, we are able to track down the vacuum persistence amplitude, a quantity of central relevance when characterizing the process~\cite{cohenmcgady}. 
Finally, we study the oscillation modes in the limit of zero fermion mass.  
We are able to observe coherent oscillations for long evolution times, finding that the energy spectrum of the time-evolved electric field and particle density display frequency peaks that can be attributed to bosonic excitations.

\textbf{\textit{Model.---}} Our starting point is a $1+1$D lattice version of quantum electrodynamics (QED) in a \textit{quantum link model} (QLM) formulation~\cite{HORN1981149,ORLAND1990647,CHANDRASEKHARAN1997455,PhysRevD.60.094502}:
\begin{equation} \label{eq_QLM}
\begin{aligned}
\hat{H}_{\mathrm{QLM}} = & -\frac{\tilde{t}}{2} \sum^{L-1}_{i=1}\left( \hat{\psi}_i^{\dagger} \hat{U}_{i, i+1} \hat{\psi}_{i+1} + \text{H.c.} \right) & \\
& +m \sum^L_{i=1} (-1)^i \hat{\psi}_i^{\dagger} \hat{\psi}_i + h \sum^{L-1}_{i=1} \hat{E}_{i, i+1},
\end{aligned}
\end{equation}
\noindent where $L$ is the number of sites between which live $L-1$ links, $\tilde{t}$ is the matter-gauge coupling strength, the matter field with rest mass $m$ on site $i$ is represented by the fermionic operators $\hat{\psi}_i^{(\dagger)}$, and the electric and gauge fields on the link between sites $i$ and $i+1$ are represented by the operators $\hat{E}_{i,i+1}$ and $\hat{U}_{i,i+1}$, respectively, which satisfy $[\hat{E}_{i,i+1},\hat{U}_{j,j+1}] = \delta_{i,j} \hat{U}_{i,i+1}$. 
We adopt here a spin-$1/2$ QLM formulation, where $\hat{E}_{i,i+1}$ and $\hat{U}_{i,i+1}$ are the spin-$1/2$ $z$ and raising operators, respectively.

When $h=0$, the $\mathrm{U}(1)$ QLM Hamiltonian~\eqref{eq_QLM} is invariant under the parity transformation, while charge conjugation is explicitly broken by the open boundary conditions. 
For $h\neq 0$, parity is also explicitly broken. 
Additionally, Hamiltonian~\eqref{eq_QLM} hosts a $\mathrm{U}(1)$ gauge symmetry with local generator $\hat{G}_i=\hat{E}_{i,i+1}-\hat{E}_{i-1,i}-\hat q_i$ with $\hat q_i=\hat{\psi}^\dagger_i\hat{\psi}_i+\big[(-1)^i-1\big]/2$, which is a discretized version of Gauss's law.  
Here, we work in the physical sector of states $\ket{\Psi}$ satisfying $\hat{G}_i \ket{\Psi} = 0,\, \forall i$, and boundary fields are set as $\hat{E}_{0,1} = \hat{E}_{L,L+1} = -\frac{1}{2} $.

\textbf{\textit{Probing false vacuum dynamics.---}} The $\mathrm{U}(1)$ QLM is implemented in our setup with ultracold $^{87}$Rb atoms in 1D optical superlattices in the presence of a tilt potential $\Delta$ ~\cite{simon2011quantum,sachdev2002mott}. 
The mapping is briefly sketched in Fig.~\ref{fig1}(b), when $U \approx \Delta \gg J$, the only hopping process allowed is $\ket{..20..} \leftrightarrow \ket{..11..}$, corresponding to the creation (annihilation) of particle pairs, further details are found in our previous works~\cite{PhysRevLett.131.073401,zhang2023observation}.

To rule out the effects of unwanted processes such as single-atom hopping, we implemented two post-selection criteria: (i) conservation of total atom number and (ii) conservation of Gauss’s law~\cite{PhysRevLett.131.050401,zohar2019removing,mildenberger2022probing,zhang2023observation}. 
Using this implementation, we can deterministically prepare the system into a Fock state, which corresponds to one of the two degenerate $\mathbb{Z}_2$ symmetric ground states of Eq.~\eqref{eq_QLM} at $h=0$ and $m\rightarrow +\infty$, also called \textit{bare} Dirac vacuum. 
These two states, in the language of the Bose--Hubbard model, are $\ket{0^+} = \ket{\ldots2020\ldots}$ and $\ket{0^-} = \ket{\ldots 0202 \ldots}$. 
As shown in Fig.~\ref{fig1}(a), they correspond to two uniform configurations of the electric field $\bra{0^\pm} \hat{E}_{i,i+1} \ket{0^\pm} = \pm 1/2,\, \forall i $. 
By adding a nonzero $h>0$ term, we can break the degeneracy between the two vacuum states, and $\ket{0^+}$ becomes a {\it false vacuum}, having finite energy density with respect to the ground state (the {\it true vacuum}) $\ket{0^-}$. 
Thereafter, by preparing the system in $\ket{0^+}$ and evolving under the $\mathrm{U}(1)$ QLM Hamiltonian with finite $m$ and $\tilde t$ (performing a \textit{quantum quench}), particle--antiparticle pairs are created. 
In our experimental setup, we can probe these processes by measuring the time evolution of on-site atom numbers with a quantum gas microscope.

The creation of particle--antiparticle pairs can be understood in this context as the effect of quantum fluctuations. 
For finite $m$ in the absence of a background field ($h=0$), creating a pair requires an energy $\Delta E\approx 2m$. 
Pairs cannot be created ``on-shell'', and the creation is suppressed for large particle mass. 
In contrast, in the presence of a background field $h>0$, since the electric field changes its sign in the region that separates the particle and the antiparticle, the energy cost of a pair on the false vacuum $\ket{0^+}$ is $\Delta E \approx 2m - h\ell$, where $\ell$ is the size of the pair (i.e., the distance between the particle and antiparticle). 
The tower of such states (called ``bubbles'') is shown in Fig.~\ref{fig1}(a). 
Because the false vacuum is at high energy, it can resonantly decay into the continuum of multi-bubble states. 
For small $h$ with respect to the mass scale --- also known as the \textit{thin wall} limit --- the decay of the false vacuum takes a long time, because the probability amplitude of creating a bubble of critical size $\ell_c \approx 2m/h \gg 1$ is exponentially small. 
In this work, we will not investigate this long-time regime, but focus on the production of small bubbles at short times. 
By tuning the parameters to a regime where $2m/h \approx 1$ (away from the thin wall limit), we can enhance the production of small bubbles, which are created ``on shell''.

Our quantum quench approach is analogous to the one proposed in previous theoretical studies of false vacuum decay in quantum spin chains~\cite{PhysRevB.60.14525,PhysRevB.104.L201106,lagnese2023detecting}, with a few differences.
The standard approach in spin chains would require performing a quench starting from a ground state at $(m=m^*, h=0)$ (inside the symmetry broken phase) to a Hamiltonian with $(m=m^*, h=h^*>0) $ with small $h^*$: this is however very resource demanding in the present setting. 
Instead, as we previously explained, we start from the bare vacuum $\ket{0^+}$, and we quench directly to $(m=m^*, h=h^*)$. 
We study two cases: $h^*>0$ comparable with $m^*$ (false vacuum), and to $h^*=0$ when investigating the contribution to particle production due to the change of the value of $m$. 
It is important to stress that the overall phenomenology of false vacuum decay --- starting from a symmetry broken state and adding a symmetry-breaking perturbation --- is expected to show some robustness. For $m^*>m_c\approx 0.655\, \tilde t/2$, the model is in a symmetry-broken phase for $h=0$: the initial state, when tuning $m$ from $\infty$ to such finite value $m^*$, is still significantly overlapped with the proper false vacuum state~\cite{Magnifico2020realtimedynamics}. 
On the other hand, for $m^*<m_c$, the ground state of the model in the absence of the $h$ field is unique, and the parity symmetry is not spontaneously broken. 
In this regime, the intuitive picture of a metastable false vacuum and a double-well potential does not apply: the energy barrier that hindered the dynamics of the $\ket{0^+}$ state is not present, and pair production is enhanced.

While this intuitive picture of an effective potential is a useful qualitative interpretation, we also provide a quantitative understanding of pair production using a perturbative effective model of our dynamics. 
By bench-marking our results against the effective model, and investigating the deviations from it, we show that our results consistently capture the effect of the external field over the pair production rate, even in regimes where higher-order perturbative corrections become non-negligible.

\begin{figure*}[tb]
    \centering
    \includegraphics[width=170mm]{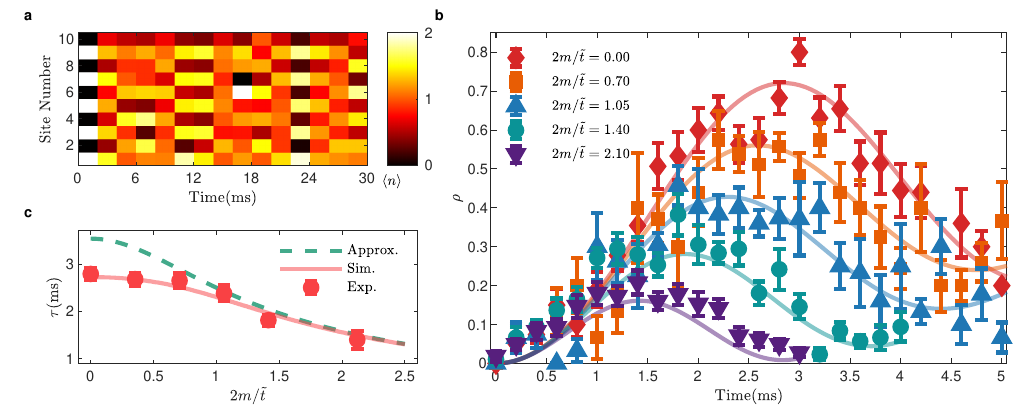}  
    \caption{\textbf{Determination of the time $\tau$ to achieve maximum particle production}. 
    (a) Single-site-resolved atom number distribution $ \langle n \rangle $ as a function of evolution time at the rest mass $m=0$ and the background field $h=0$. 
    (b) Averaged particle density $\rho$ as a function of evolution time at different rest masses.
    We extract the peak times from the experimental data with Gaussian fitting. 
    Solid lines represent numerical simulations carried out with experimental parameters in Hamiltonian Eq.~\eqref{eq_QLM}.  
    (c) The extracted peak time $\tau$ as a function of the rest mass $m$. 
    As the rest mass $m/\tilde{t}$ increases, $\tau$ converges towards the relationship of $\tau =\pi/\sqrt{\tilde t^2+4m^2}$. 
    The solid line represents the results of numerical simulations, and the green dashed line in (c) signifies the prediction of the two-level model Eq.~\eqref{eq:twolevelrho}. 
    Error bars denote the standard error of the mean (s.e.m.). }
    \label{fig2}
\end{figure*}

\begin{figure*}[tb]
    \centering
    \includegraphics[width=170mm]{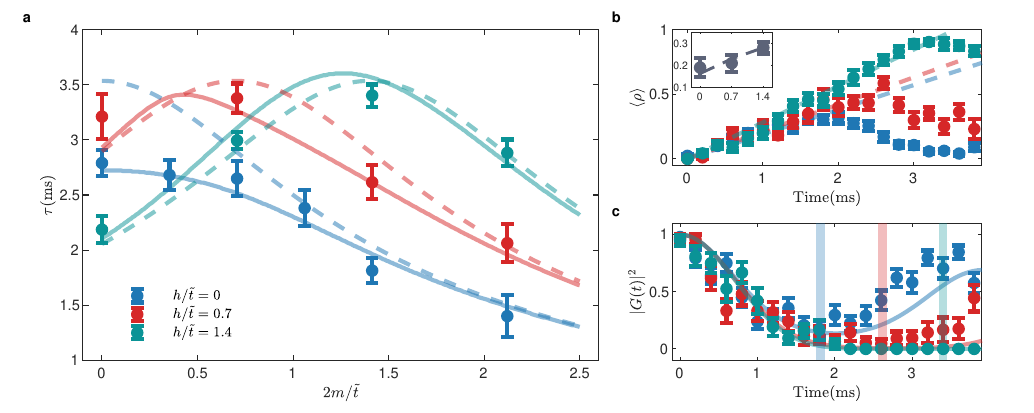}
    \caption{\textbf{Decay rate of false vacuum with external field}. 
    (a) Peak time $\tau$ as a function of $2m/\tilde{t}$ for different external field strengths $h/\tilde{t} = \{0,0.7,1.4\}$. 
    The quantity exhibits a peak at $2m\approx h$, where pairs can be produced resonantly. 
    Solid lines depict the exact numerics in Hamiltonian Eq.~\eqref{eq_QLM} at $2m/\tilde{t} = 1.4, h/\tilde{t} =\{0,0.7,1.4\} $. 
    Dashed lines represent the results of the two-level approximation.  
    (b) External-field-dependent dynamics of particle density in the initial increasing process with linear fitting at $2m/\tilde{t} = 1.4, h/\tilde{t} =\{0,0.7,1.4\} $. 
    The dots denote the experimental data with errors, and the dashed lines indicate the linear fitting of the particle density. 
    Inset: The particle density growth rate, $\dot{\rho}$, positively correlates with the external field. The fit was performed over the time interval from $0$ to $\tau$. 
    The dashed line is obtained from the two-level approximation $\rho_{\text{max}}/\tau$.
    (c) The time evolution of the vacuum persistence amplitude at $2m/\tilde{t} = 1.4, h/\tilde{t} =\{0,0.7,1.4\} $. 
    Translucent vertical lines in (c) mark the peak time $\tau$.  
    Error bars denote the s.e.m. in (a-c).} 
    \label{fig3}
\end{figure*}

\begin{figure}[tb]
    \centering
    \includegraphics[width=80mm]{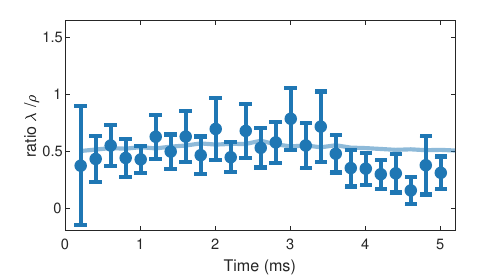}
    \caption{\textbf{Vacuum persistence amplitude versus particle production}. 
    Ratio $\lambda/\rho$ as a function of time for $2m/\tilde{t}=1.4, h/\tilde{t}=0.7$. 
    Until timescales of order of 4 ms, the ratio is approximately constant within error bars. 
    Error bars denote the s.e.m.. 
    The blue line is the result of a two-domain wall approximation.} 
    \label{figvpa}
\end{figure}

\begin{figure*}[tb]
    \centering
    \includegraphics[width=170mm]{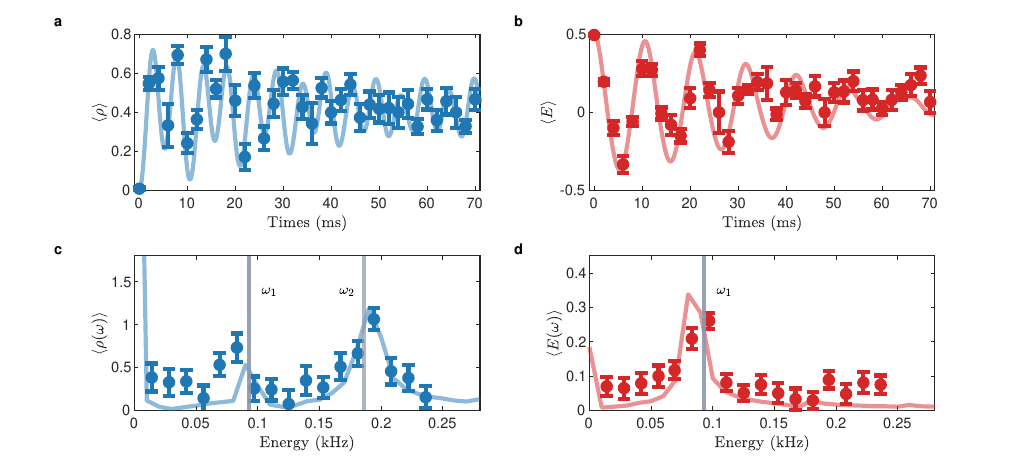}
    \caption{\textbf{Energy spectrum for the massless case}. 
    (a) Real-time particle density dynamics for $m=0,h=0$. 
    (b) The real-time dynamics of $\langle E \rangle$ for the same case. 
    (c-d) Fourier amplitudes of the real-time oscillations of $\braket{\rho}$ and $\braket{E}$. 
    The grey vertical lines represent the first two energy gaps, $\omega_1$ and $\omega_2 \approx 2\omega_1$, of the Hamiltonian Eq.~\eqref{eq_QLM} by exact diagonalization, respectively. 
    Error bars denote the s.e.m. in (a-b). 
    The error bars in (c-d) were derived using error propagation for the Fourier transformation~\cite{582534}.}
    \label{fig4}
\end{figure*}

\textbf{\textit{Particle production without and with external field.---}} To better understand the particle--antiparticle pair production process from the vacuum, we focus on determining the peak time $\tau$ required to achieve maximum particle production. 
First, we fix the external field $h=0$ and start in a $\mathbb{Z}_2$ symmetry-broken vacuum state with an atom occupation in the Fock state $\ket{\ldots2020\ldots}$, as shown in Fig.~\ref{fig1}(b). 
In the resonant condition $m=0$, we observe a slowly decaying oscillation from the measured site-resolved atom number distribution, with a frequency correlated to the matter-gauge coupling strength $\tilde{t}=75(1)$ Hz, as shown in Fig.~\ref{fig2}(a). This regime is akin to that explored in Rydberg atom experiments (where only $h=0$ was considered)~\cite{Bernien2017}.

In Fig.~\ref{fig2}(b), we plot the averaged particle number density $\rho(t) = \sum_i |\hat q_i|/ L$ as a function of evolution time for various values of the masses for the system size $L=10$. 
We then determine $\tau$ as the time when the particle number density $\rho(t)$ reaches its first local maximum. The extracted results are shown in Fig.~\ref{fig2}(c).
In Fig.~\ref{fig2}(b), we observe that the first peak of the particle density $\rho$ becomes lower and appears early as the particle mass $m$ increases, which means that the pair production is suppressed by the energy barrier due to the rest mass $m$.

The local pair production after the quench can be explained with a two-level model of decoupled oscillations, which is expected to be exact in the large-mass $m \gg \tilde{t}$ regime (namely, when the density of pairs is sufficiently small that the recombination of two neighboring pairs through the annihilation of a particle and an antiparticle is perturbatively suppressed). 
The model is described by the effective Hamiltonian
\begin{equation}
\label{eq:twolevelham}
    \hat{H}_\text{eff} = 
    \begin{pmatrix}
       0 & \tilde{t}/2\\
       \tilde{t}/2 & 2m 
    \end{pmatrix},
\end{equation}
\noindent which describes oscillations between the zero-particle state and the particle--antiparticle state. 
The density of pairs is then nothing but the probability of being in the particle--antiparticle state, obtained by diagonalizing Eq.~\eqref{eq:twolevelham} and evolving the zero-particle initial state, leading to 
\begin{equation}
\label{eq:twolevelrho}
\rho(t)= \frac{\tilde t^2}{4m^2+\tilde t^2} \sin^2 \left(\frac{1}{2}\sqrt{\tilde t ^2+4m^2} \; t\right).
\end{equation}

The predicted peak time $\tau$ is then $\tau= \pi/\sqrt{\tilde t ^2+4m^2} $. 
In Fig.~\ref{fig2}(c), the peak time extracted from experimental data is compared with the one predicted from the two-level model, reporting good agreement for $2m>\tilde{t}$.  
For small values of the mass, the data shows significant deviations from the two-level approximation: In this regime, pairs are not diluted, so the process of annihilation that leads to the formation of larger bubbles plays a role in the evolution.

We now focus on the case in the presence of a finite external field $h$, which can be realized by changing the depth of the staggered superlattice potential in our experiment~\cite{zhang2023observation}. 
The initial state in this scenario is the false vacuum $\ket{0^+}$, and we observe the time evolution dynamics of this state under the $\mathrm{U}(1)$ QLM Hamiltonian~\eqref{eq_QLM}. 
We plot the extracted peak time $\tau$ as a function of the rest mass at different external fields $h$ in Fig.~\ref{fig3}(a). 
We observe that a large background field $h$ can suppress vacuum fluctuations and prevent the conversion of energy in the vacuum to particles in both small- and large-mass $m$ ranges. 
As shown in Fig.~\ref{fig3}(a), the two-level model can be used again to interpret the results by substituting $2m$ with $2m - h$ in Eqs.~\eqref{eq:twolevelham} and~\eqref{eq:twolevelrho}. 
As long as $2m-h$ is sufficiently big, in the short timescale, further recombination processes of particle and antiparticles between neighboring sites are perturbatively suppressed, and the dynamics remain approximately local. 
Thus local production is governed by the competition of the diagonal and off-diagonal terms in Eq.~\eqref{eq:twolevelham}. 
In the near-resonant range, where $2m \approx h$, the energy of the background field approximately matches the energy required to create pairs of particles and antiparticles. 
Indeed, here we find the maximum value of $\tau$, and the largest deviations from the two-level model (unless $h$ is sufficiently large), demonstrating the partial recombination of particle--antiparticle pairs.

We then investigate the effect of the external field on the rate of particle production $\dot\rho$. 
Its magnitude is extrapolated from the experimental data through a linear fit from $t=0$ to $t=\tau$, as shown in Fig.~\ref{fig3}(b). 
The production rate extracted from the fit is well-captured by the two-level model as $\rho_\text{max}/\tau\approx\tilde t^2/\pi\sqrt{(2m-h)^2+\tilde t^2}$. 
Note that, in this regime, the pair production can arise as a low-order process in perturbation theory, so the well-known non-perturbative expression of the Schwinger formula does not apply here \cite{Magnifico2020realtimedynamics}.

\textbf{\textit{Vacuum persistence amplitude.---}} Another important quantity to qualify the decay of the unstable vacuum is the vacuum persistence amplitude $ G(t) = \langle \psi(0) | \psi(t) \rangle $, which measures the deviation from the vacuum state. 
Thanks to single-site resolution, our experiment can address this. 
We plot $|G(t)|^2$ as a function of evolution time at various external fields in Fig.~\ref{fig3}(c). 
We observe that for $h/\tilde{t}=0,0.7$, $|G(t)|^2$ reaches its minimum at the time $t=\tau$, while for $h/\tilde{t}=1.4$, $|G(t)|^2$ drops rapidly to a very small value, preventing us from extracting the time of the minimum from the data.
These results are consistent with those shown in Fig.~\ref{fig3}(a), indicating that in the intermediate regime of $2m \approx h$, $|G(t)|^2$ can drop rapidly to zero as the energy required to generate the particle--antiparticle pair is approximately resonant with the energy of the background field. 
This can be understood, once again, using the two-level approximation in the dilute limit. 
Within this approximation, the vacuum persistence amplitude takes the form $|G(t)|^2\approx (1-\rho(t))^{L/2}$ leading to the result that $\rho$ and $|G(t)|^2$ reach their extreme value simultaneously. Moreover, if $\rho(t)$ is small we get $|G(t)|^2\approx \exp(-L \rho(t)/2)$. 
Since Schwinger proposed his interpretation for the phenomenon of pair production, $\rho$ and $|G(t)|^2$, and more precisely $ \lambda(t) = -L^{-1}\log(|G(t)|^2)$\cite{Muschik_2017}, became two interesting quantities to monitor in reciprocal relation~\cite{cohenmcgady}. 
The direct relation $\rho(t) \propto \lambda(t)$ that we derive within the approximation of decoupled oscillators is expected to hold more generally, also in the continuum, when $\rho$ is small and the process is led by local independent emission~\cite{itzyksonzuber} (see also~\cite{GELIS20161,cohenmcgady} for a detailed discussion). 
In Fig.~\ref{figvpa}, we plot the ratio $\lambda(t)/\rho(t)$ as a function of time during the particle production process. 
We observe that, up to timescales of the order of 4 ms, the ratio is constant within error bars. 
The observation is also in agreement with the effective model of decoupled oscillators (blue line).

\textit{\textbf{Diabatic spectroscopy through long-time dynamics.---}} We further analyze the oscillation modes when evolving from the bare Dirac vacuum $\ket{0^+}$ with $m=h=0$, and we characterize the particle content by analyzing the amplitude spectrum of the signal. 
Note that in this massless regime, the ground state of the system does not break the $\mathbb{Z}_2$ charge conjugation symmetry: the two vacua $\ket{0^\pm}$ are not low-energy states, but rather lie in the middle of the spectrum. 
The observed phenomenology makes a connection with the phenomenon of \textit{quantum many-body scars} that were observed in the PXP model~\cite{Turner2018,PhysRevResearch.5.023010}, to which our model can also be mapped.

In Fig.~\ref{fig4}(a,b), we show the measured values of the averaged particle density $\rho$ and the averaged electric field strength $ E  = \sum_i E^z_{i,i+1} / (L-1)$, respectively, as a function of the evolution time after the quench. 
In this regime, many particles can be produced (note the large values of the maximum density $\braket{\rho}\approx 0.7$), and the recombination and annihilation process can lead to the inversion of the string, such that $E$ oscillates between positive and negative values. 
We can evolve the system for sufficiently long times to observe $6$ full periods of oscillations for the electric field. 
These long-lived oscillations have been attributed to the presence of special scar states in the many-body spectrum.

The corresponding Fourier-transformed data are $ \rho(\omega), E(\omega)$ in Fig.~\ref{fig4}(c,d)~\cite{Kormos2017,Tan2021}. 
Two peaks with frequencies $\omega_1= 0.08(1) $ kHz and $2\omega_1$ are clearly visible in the fermion density. 
A peak at $\omega_1$ is also visible in the Fourier transform of the electric field. 
Since the initial state has high energy, the frequency $\omega_1$ associated with the long-lived string-inversion oscillations cannot be easily attributed to the low-lying excitations. 
However, various works have shown that approximate quasiparticle treatments can give useful insights into their phenomenology \cite{Chandran2023}.

In this context, the spin-$1/2$ QLM displays very similar phenomenology as the massless Schwinger model, which can be mapped to a free scalar bosonic field theory \cite{COLEMAN1976239}. 
The oscillatory modes of the electric field can be attributed to these bosonic excitations \cite{PhysRevX.10.021041}. 
In the Schwinger model, the oscillations are persistent because the bosons are stable non-interacting particles. 
For the massless QLM, on the other hand, the spectrum cannot be exactly expressed in terms of stable quasiparticles. 
Nevertheless, one can observe long-lived oscillations, which may be the remnant of the persistent string inversion in the field theory.

\textbf{\textit{Conclusion and outlook.---}}
We have presented a quantitative experimental study of false vacuum decay in a $\mathrm{U}(1)$ lattice gauge theory with ultracold bosons in optical lattices. 
The particle production rate during the decay process was found to be suppressed either by an increase in the rest mass $m$ or by the dominance of the external field $h$.
Furthermore, as the fermion mass approaches infinity, the production rate decreases inversely proportional to mass. 
By tracking the vacuum persistence amplitude, we have revealed the direct relationship between pair production and vacuum decay, i.e., the production rate is approximately proportional to the vacuum decay rate.
Moreover, the oscillation modes in the limit of zero fermion mass have been extracted and the frequency peaks in the spectrum are compatible with bosonic excitations and string inversion phenomenology.  
Our observations agree well with the numerical benchmark of exact diagonalization~\cite{10.21468/SciPostPhys.7.2.020}, demonstrating the potential for neutral-atom quantum simulators to quantitatively study real-time dynamics of lattice gauge theories. 
The tunable external field capability developed in this work can be explored for exotic phenomena such as dynamical quantum phase transitions~\cite{PhysRevLett.122.250401,PhysRevLett.122.050403,osborne2024mesonmasssetsonset}, string breaking~\cite{PhysRevLett.109.175302,PhysRevLett.111.201601}, Hilbert space fragmentation \cite{Desaules2024ergodicitybreaking}, 
and meson scattering~\cite{surace2021scattering,su2024coldatomparticlecollider}. 
The methods presented to study the oscillation modes can also be extended to other spectroscopy works, e.g., detecting the low-energy excitations in true (false) vacuum~\cite{lagnese2023detecting}. 
Possible extensions of our work include systems in $2+1$D~\cite{balducci2022localization} and other gauge theories~\cite{PhysRevD.106.025008}.

\begin{acknowledgments}
This work was supported by NNSFC grant 12125409, Anhui Provincial Major Science and Technology Project 202103a13010005, Innovation Program for Quantum Science and Technology 2021ZD0302000. 
G.L.~was supported by P1-0044 program of the Slovenian Research Agency, the QuantERA grants QuSiED and T-NiSQ by MVZI, QuantERA II JTC 2021, and ERC StG 2022 project DrumS, Grant Agreement 101077265. 
F.M.S.~acknowledges support provided by the U.S.\ Department of Energy Office of Science, Office of Advanced Scientific Computing Research, (DE-SC0020290); DOE National Quantum Information Science Research Centers, Quantum Systems Accelerator; and by Amazon Web Services, AWS Quantum Program. 
W.-Y.Z.~acknowledges support from the Postdoctoral Fellowship Program of CPSF under Grant Number GZC20241659.
J.C.H.~acknowledges support from the Max Planck Society. 
M.D.~was partly supported by the QUANTERA DYNAMITE PCI2022-132919, by the EU-Flagship programme Pasquans2, by the PNRR MUR project PE0000023-NQSTI,  and by the PRIN programme (project CoQuS).

\end{acknowledgments}

\bibliography{fvd_ref_nourl}
\bibliographystyle{Science}

\onecolumngrid
\vspace*{0.5cm}
\newpage
\begin{center}
    \textbf{METHODS AND SUPPLEMENTARY MATERIALS}
\end{center}
\vspace*{0.5cm}

\twocolumngrid
\incltocpage
\tableofcontents
\appendix
\setcounter{secnumdepth}{2}

\twocolumngrid
\setcounter{equation}{0}
\setcounter{figure}{0}
\makeatletter
\makeatother
\renewcommand{\theequation}{S\arabic{equation}}
\renewcommand{\figurename}{Extended Data Fig.}
\renewcommand{\thefigure}{\arabic{figure}}
\renewcommand{\thetable}{S\arabic{table}}

\section{Mapping between U(1) gauge theory model and Bose-Hubbard model}

In our experiment, the tilted Bose-Hubbard model (BHM) with a staggered potential and open boundary conditions, described by the Hamiltonian
\begin{equation} 
\label{BH1}
\begin{split}
    \hat{H}_{\textrm{BHM}} =& -J \sum^{L-1}_{i=1} \left( \hat{b}^{\dagger}_i \hat{b}_{i+1} + 
    \hat{b}^{\dagger}_{i+1} \hat{b}_{i}  \right) \\ & + \frac{U}{2}\sum^L_{i=1} \hat{n}_i ( \hat{n}_i -1 ) + \sum^L_{i=1} \epsilon_i \hat{n}_i ,
\end{split}
\end{equation}
\noindent where $J$ is the hopping amplitude, $U$ is the interaction interaction and $L$ is the number of lattice sites. 
The energy offset $\epsilon_i = (-1)^i \delta/2 + i\Delta $ consists of a staggered superlattice potential $\delta$, and a liner tilt potential $\Delta$. 
In the near-resonant condition $U \approx \Delta \gg J$, and $\delta = 0$, the mainly allowed hopping process is $\ket{20} \leftrightarrow \ket{11}$. 
So in the first order approximation, the Bose-Hubbard Hamiltonian \ref{BH1} maps to the U(1) gauge theory in the main text, where the configuration $20$ maps to $\uparrow$ and the other configurations $(11,12,02,01)$ maps to $\downarrow$ \cite{zhang2023observation}.

\begin{figure}[!htb]
    \centering
    \includegraphics[width=80mm]{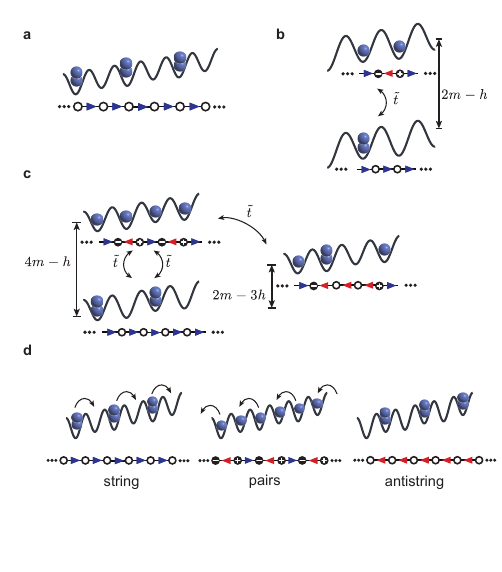}
    \caption{\textbf{Sketch of pair recombination and string inversion.} 
    (a)The false vacuum state and its configurations in BHM. 
    (b)The process of pair production and annihilation in a two-site model. 
    (c)The pair recombination process. 
    If two adjacent pairs of particles are produced, they could recombine and leave a pair at a longer distance at the end of the string. 
    (d)Schematic of string inversion at $m=h=0$. 
    The initial string $\ket{...202020...}$ evolves to pairs states and finally comes to the antistring state by the pair recombination process.   }
    \label{sup_fig_sketch}
\end{figure}

Under this mapping, the $\ket{20} \leftrightarrow \ket{11}$ hopping in BHM corresponds to the creation and annihilation of particle pairs, which results in the inversion of the gauge field. 
In the small mass regime $m/\tilde{t} \ll 1$, there will be more than one pair created simultaneously, and the pair annihilation can also occur between two neighboring pairs, see Fig.~\ref{sup_fig_sketch}. 
The pair recombination process leaves the particle-antiparticle pair at the end of the string and expands the length of the string. 
But for the large mass limit $m/\tilde{t} \gg 1$, the process is suppressed by $2m$, and the local creation (annihilation) process dominates, which can be captured by the two-level model.

\section{Numerical simulations}

The numerical results presented in the main text and this supplementary material are based on exact diagonalization (ED).

\subsection{Quench dynamics of Bose-Hubbard model and QLM}

To numerically verify the mapping between U(1) gauge theory and Bose-Hubbard models, we set the Hubbard parameters at the resonance condition $\Delta=U=800$Hz and $J=50$Hz for convenience. 
The QLM parameters $\tilde{t} = 50\sqrt{2}$ Hz as deduced in \cite{zhang2023observation}. Then, we compute the quench dynamics of both the Bose-Hubbard model and QLM by initializing the state to $|2020...\rangle$ and $|\text{vacuum}\rangle$, respectively, with system size $L = 10$.

\begin{figure}[tb]
    \centering
    \includegraphics[width=80mm]{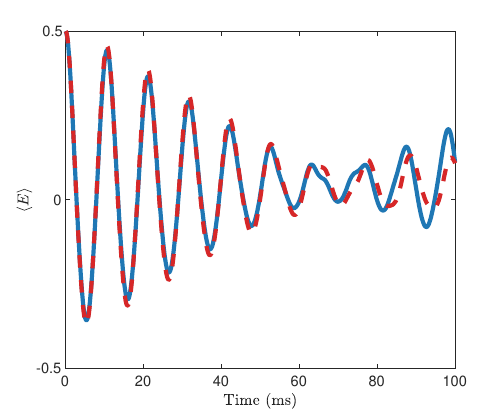}
    \caption{\textbf{Time evolution of electric field strength $\braket{E(t)}$.} 
    The numerical results of $\braket{E(t)}$ from the Bose-Hubbard model (blue line) and QLM (red dashed line) for the resonant condition $U=\Delta$ and $m=0$, respectively. }
    \label{sup_fig1}
\end{figure}

The time evolution of particle density $\braket{\rho(t)}$ and electric field strength $\braket{E(t)}$ is shown in Fig.~\ref{sup_fig1}, which were experimentally measured in the main text. 
We find good agreement between the Bose-Hubbard model and QLM after the quench until $tJ \approx 18 $, and the difference mainly comes from states beyond the constraint Hilbert space. 
We also study the quality of this approximation in our experimental parameters while the external field $h$ is present. 
In the time region of experiment $t<10$ ms, the behavior of $\braket{\rho(t)}$ and $\braket{E(t)}$ agree well with the numerical results.

\begin{figure*}[!htb]
    \centering
    \includegraphics[width=\linewidth]{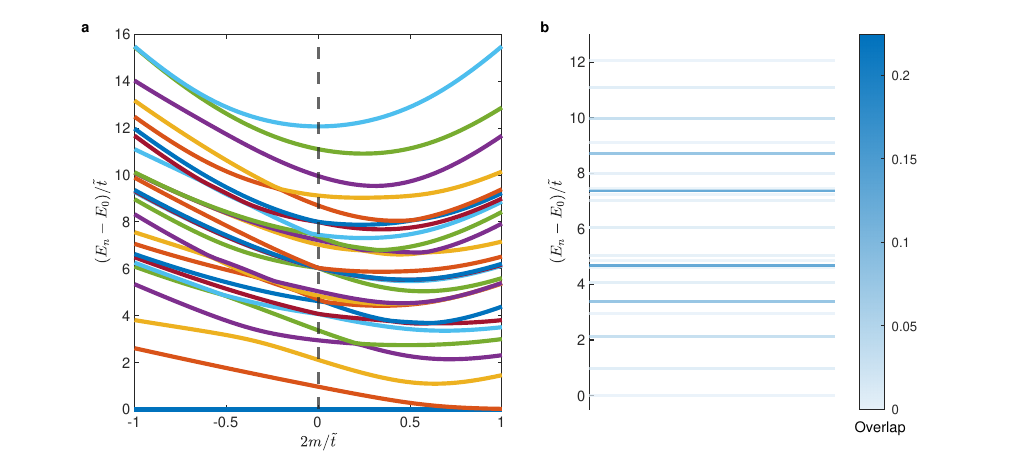}
    \caption{\textbf{Energy level of QLM}. 
    (a) The energy spectrum of QLM for the case $h=0$. (b) The energy level and overlap between eigenstates and the vacuum state. 
    Color represents the overlap $|C_n|^2$ between the vacuum state and n-th eigenstate $\ket{n}$. } 
    \label{fig_gap}
\end{figure*}

\subsection{Vacuum persistence amplitude}

The vacuum persistence amplitude 
\begin{equation}
    \mathcal{G}(t) = \braket{ \text{vac}|e^{-i\hat{H}t} | \text{vac} },
\end{equation}
\noindent implies the instability of the vacuum due to pair production and is an important quantity to illustrate the dynamics of the Schwinger model. 
In the continuum limit, the particle density can be obtained by the vacuum decay $\rho(t) = \lambda(t)$ with $\lambda(t)=-N^{-1} \log (|\mathcal{G}(t)|^2) $. 
The quantity is also important in the context of dynamical quantum phase transition where $|\mathcal{G}(t)|^2$ and $\lambda(t)$ are known as the Loschmidt echo and rate function, respectively.
In the main text, we work in a discrete lattice with truncation in the gauge field, so there is a difference between the measured $\rho$ and $\lambda(t)$. 
However, their behaviors are similar, and they can be roughly explained in the two domain wall model. 
We restrict the Hilbert space in states with no pair $\ket{\psi_{0p}}$ and with only one pair $\ket{\psi_{1p}}$,
\begin{equation}
    \ket{\psi(t)} = c_{0}(t) \ket{\psi_{0p}} + c_{1}(t) \ket{\psi_{1p}},
\end{equation}
\noindent where the coefficients $|c_{0}|^2 = 1-n(t)$ and $|c_{0}|^2 = n(t)$ and the vacuum persistence amplitude and particle density can be expressed by $n$,
\begin{equation}
    \lambda(t) = -L^{-1}\log(1-n), \quad  \rho(t)  = -L^{-1}2n,
\end{equation}
\noindent and in the small $n$ limit, the ratio $\lambda(t)/\rho(t) \approx \frac{1}{2}+\frac{n(t)}{4} $ is nearly constant.

\subsection{Excitation spectrum of QLM}

As mentioned in the main text, we extract the excitation spectrum of QLM with Fourier transformation. 
The initial state $\ket{\textit{vacuum}}$ can be written in the eigenstate $\ket{n}$ basis $\ket{\textit{vacuum}} = \sum_n C_n \ket{n}$ and the time evolution of any observables $\braket{M(t)}$ is given by:
\begin{equation} \label{Eq_Fourier}
    \braket{M(t)} = \sum_{nn'} C_{n}C_{n'}^* \exp^{-i(E_n-E_{n'})} \bra{n'}M\ket{n},
\end{equation}
\noindent where $E_n$ is the energy of n-th eigenstate $\ket{n}$. 
Therefore, by Fourier transformation, we can extract the energy difference $\Delta E_{nn'} = (E_n-E_{n'})$ with the amplitude $C_{n}C_{n'}^*\bra{n'}M\ket{n}$.

To better understand our spectroscopy results, we calculate all eigenenergy and eigenstates of QLM Hamiltonian using exact diagonalization at $L=10$ and $m=h=0$, see Fig.~\ref{fig_gap}. 
The initial vacuum state is not a low energy state but has a large coefficient $|C_n|^2$ with eigenstates in the middle of the spectrum, e.g., $\{n\}=\{5,9,13,19,23\}$.  
We observe that the energy level spacing is relatively constant, especially for eigenstates largely overlapping with the vacuum state. 
Their energy difference $\Delta E \approx 1.3 \Delta E_{2,1}$.
Thus, the coupling and energy difference among these eigenstates are significant in the Fourier spectrum, see Eq.~\ref{Eq_Fourier}, and peaks extracted from the spectrum are quantitatively close to the energy gap of the massless QLM.

\begin{figure*}[!htb] \label{fig_overlap}
    \centering
    \includegraphics[width=80mm]{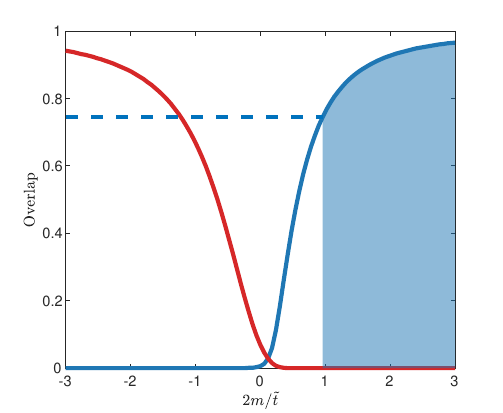}
    \caption{\textbf{Overlap between ground state and product states}. 
    Overlap between ground state at various mass $2m/\tilde{t}$ and two product states $\ket{20...}$ (blue line) and $\ket{11...}$ (red line). 
    While the mass goes across $2m/\tilde{t} = 1$, the overlap comes relatively large.} 
    \label{fig_overlap}
\end{figure*}

\section{Experimental sequence}

\subsection{Initial state preparation}

Our experiment starts from a two-dimensional (2D) Bose-Einstein condensate of $^{87}$Rb atoms in the $\ket{F=1,m_{F}=-1}$ state, which has been described in our previous work. 
We then prepare the multiple copies of one-dimensional near-unity-filled Mott insulators along the $y$ direction utilizing a staggered immersion cooling method.  
By tuning the lattice potential and the relative phase of the long lattice, we merge the atoms in the double wells along the $y$ direction according to the time sequence illustrated in Figure S3 and finally prepare copies of initial Fock state $\ket{\psi_0} = \ket{2020...}$. 
According to the mapping between the U(1) gauge theory model and the Bose-Hubbard model, the initial state is one of the vacuum states and is one of the ground states for $m \rightarrow +\infty$. 
As for preparing the ground state of finite $m$, the adiabatic ramp from $m \rightarrow +\infty$ is the most used method but is limited by efficiency. 
Fortunately, we numerically calculate the overlap between the ground state of finite $m$ and the vacuum state, see Fig.~\ref{fig_overlap}. 
We observe that while $m$ crosses the critical point $2m/\tilde{t}>0.655$, the overlap is significantly large and increases to $90\%$ for $2m/\tilde{t}>2$.

\subsection{Quench and evolution}

After the false vacuum state has been manipulated, the gradient field is ramped to its final value in the first instance. 
Next, we quench system parameters to the desired value within 0.2 ms and then allow the system to evolve freely. 
While the lattice depth is ramped down, atoms start to evolve, so the quenching process should be as quick as possible without heating atoms. 
Following the time evolution, we abruptly increase the lattice depth back to freeze the atom and switch off the gradient field.

\begin{figure*}[htb]
    \centering
    \includegraphics[width=\linewidth]{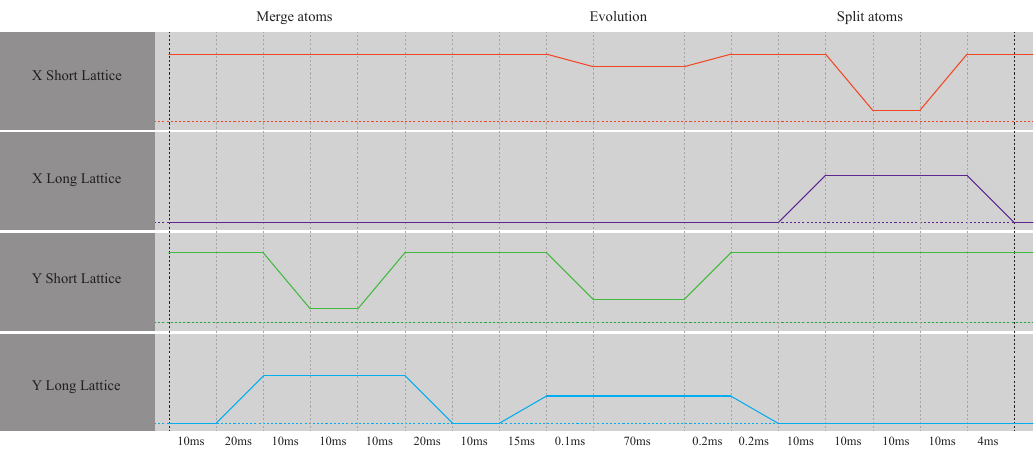}
    \caption{Experimental sequences. 
    Sequence for merging atoms from $|\dots111111\dots\rangle$ to $|\dots202020\dots\rangle$, the evolution of false vacuum decay, and the splitting process for state readout.}
    \label{sup_fig3}
\end{figure*}

\end{document}